\documentclass[12pt,english]{article}
\usepackage[LGR,T1]{fontenc}
\usepackage[latin1]{inputenc}
\usepackage{geometry}
\usepackage{amsmath}
\usepackage{amssymb}
\usepackage{graphicx}

\makeatletter

\DeclareRobustCommand{\greektext}{%
  \fontencoding{LGR}\selectfont\def\encodingdefault{LGR}}
\DeclareRobustCommand{\textgreek}[1]{\leavevmode{\greektext #1}}
\ProvideTextCommand{\~}{LGR}[1]{\char126#1}

\@ifundefined{date}{}{\date{}}

\usepackage{units}\usepackage{amstext}\usepackage{dsfont}\usepackage{setspace}\usepackage{mathrsfs}\usepackage{longtable}\usepackage{times}\usepackage{ulem}\usepackage{babel}

\newcommand{\suppress}[1]{}

\newlength\wvtextpercent
\newbox\strikebox
\def\strike#1{\setbox\strikebox \hbox{<#1>}\hbox{\raise0.5ex\hbox to 0pt{\vrule height 0.4pt width \wd\strikebox\hss}\copy\strikebox}}

\makeatother

\usepackage{babel}
\begin{document}


\linespread{1.5}
\title{On the time continuous evolution of the universe if time is discrete
and irreversible in nature }
\author{\textbf{Roland Riek} \\
 \textbf{Laboratory of Physical Chemistry, ETH Zurich, Switzerland}}

\maketitle
e-mail: roland.riek@phys.chem.ethz.ch\\
\\
This is the version of the article before peer review or editing, as submitted by an author to {insert name of Journal}. IOP Publishing Ltd is not responsible for any errors or omissions in this version of the manuscript or any version derived from it. The Version of Record is available online at doi = {10.1088/1742-6596/1275/1/012064} in the journal of physics conference series  1275 (1) 2019 

\section{Abstract}

The time evolution of the universe is usually mathematically described
under a continuous time and thus time reversible. Here, the consequences
of studying the evolution of a homogenous isotropic universe by time
continuous reversible physics are studied if time is actually discrete
and irreversible in nature. The discrete dynamical time concept of
Lee and its continuous time limit to the continuous time case is applied
to the Newtonian limit of the general relativity theory. By doing
so, the cosmic constant as well as the inflation of the universe arise
and are predicted quantitatively well by assuming the smallest time
step to be the Planck time and by using the current size of the universe.

\textbf{\large{}\newpage}{\large\par}

\section{Introduction}

Because experimentally measured time is composed of an array of events,
which can be exemplified in physics only by an energy-consuming clock
measurement, and thus can not be measured continuously because of
the uncertainty principle between energy and time ($\text{\textgreek{D}E \textgreek{D}t > h/2}$
with h being the Planck constant) time may be regarded discrete in
nature (see for example {[}1-10{]}). It is however not a very popular
concept, because the continuity of time and space enables powerful
mathematical tools. Nonetheless, several approaches on discrete time
physics have been elaborated on ({[}6{]}- {[}12{]}). Here, the evolution
of a homogenous isotropic universe is explored under a discrete dynamical
time and its consequences are discussed if time continuous reversible
physics is applied to an universe that expands in time steps. The
discrete dynamical time concept of Lee {[}6{]} is thereby used and
its limit to the continuous time case is applied to the Newtonian
limit of the general relativity theory {[}16{]}. This approach is
therefore not a quantum mechanical-based or a quantum-loop gravity-based
theory {[}12{]}, while it circumvents principle issues on discrete
time physics such as the Lorentz transformation, but yields an alternate
explanation on the apparent inflation of the universe and the cosmic
constant. After a short summary on several definitions and laws in
cosmology (3.1) and the introduction of a discrete dynamical time
(3.2), the request on time reversibility to Newton's equation is discussed
by introducing a scaling of time (3.2). In 3.3 and 3.4, this concept
is applied to the gravitational force yielding the cosmic constant
as ad hoc introduced by Einstein {[}13{]}. In 3.5 the evolution of
the universe is described yielding an evolutionary phase of the universe
as postulated by Guth {[}14{]} and in 3.6 the apparent acceleration
of the universe is discussed. 

\section{Theory}

\subsection{The standard universe evolution under a continuous time}

In the classical Newtonian description of gravity, the gravitational
force $\mathbf{F}_{G}(\mathbf{r})$ on a symbolic point particle of
mass $m$ at radius $r$ from the center of the mass $M$ of a point-like
particle or a spherically symmetric object (with size smaller than
$r$) is given by 
\begin{equation}
\mathbf{F}_{G}(\mathbf{r})=m\,\ddot{\mathbf{r}}=-\frac{4G}{3\pi}\rho\,\mathbf{r}\,m=-\frac{G\,M\,m}{\mathbf{r}^{2}}\:\frac{\mathbf{r}}{|\mathbf{r}|}
\end{equation}

with $G$ the gravitational constant and $\mathbf{r}$ the vector
with length $r$ from the origin of reference at the center of mass
M, which is considered here the mass of the universe, to the coordinates
of the point particle (please note, vectors are written in bold) yielding
for $\mathbf{r}=\left(\begin{array}{c}
r^{x}\\
r^{y}\\
r^{z}
\end{array}\right)=\left(\begin{array}{c}
r^{x}\\
r^{x}\\
r^{x}
\end{array}\right)$ with $r_{x}=\frac{r}{\sqrt{3}}$ . With the theorem of Gauss and
the gravitational field potential $\text{\textgreek{F}}(\mathbf{r})=-\frac{G\,M\,}{|\mathbf{r}|}$
the following Poisson equation can be obtained: 

\begin{equation}
\nabla^{2}\text{\textgreek{F}}(\mathbf{r})=\varDelta\text{\textgreek{F}}(\mathbf{r})=4\,\pi\,G\,\rho(\mathbf{r})
\end{equation}
with $\rho(\mathbf{r})$ being the mass density of the system with
mass $M$. Thus, a Newtonian gravitational field can be derived from
a scalar potential $\text{\textgreek{F}}$ that follows the Poisson
equation.

In contrast, the Poisson equation derived from the Einstein field
equations of the theory of general relativity at the Newtonian limit
{[}15{]} is

\begin{equation}
\varDelta\text{\textgreek{F}}(\mathbf{r})=4\,\pi\,G\,\rho(\mathbf{r})-\text{\textgreek{L}}
\end{equation}
with $\text{\textgreek{L}}$ defined as the cosmic constant, which
is constant by definition. Nowakowski et al. {[}16{]} showed using
the Schwarzschild solution that the potential for a point-like and/or
spherical symmetric object with mass M at the Newtonian limit is

\begin{equation}
\text{\textgreek{F}}(\mathbf{r})=-\frac{G\:M}{|\mathbf{r}|}-\frac{1}{6}\text{\textgreek{L}}\,\mathbf{r}^{2}
\end{equation}

This yields the following gravitational force acting on the point
particle of mass $m$ at radius $r$:

\begin{equation}
\mathbf{F}_{G}(\mathbf{r})=m\,\ddot{\mathbf{r}}=-\frac{G\,M\,m}{\mathbf{r}^{2}}\:\frac{\mathbf{r}}{|\mathbf{r}|}+\frac{m}{3}\text{\textgreek{L}}\,\mathbf{r}
\end{equation}

Another equation of interest here is the space time metric in case
of a flat space time geometry (restricted to a flat space time geometry
for simplicity reasons) given by

\begin{equation}
dx^{2}=(cdt)^{2}-R(t)^{2}dr^{2}
\end{equation}

with $dx$ the space time interval, which is the distance between
two events, $c$the speed of light, and $R(t)$ the scaling factor.
It describes the radial evolution of an isotropic homogenous universe
under a continuous time $t$under the theory of general relativity
{[}13, 15{]}. We further mention the Hubble law {[}21{]} with the
time-dependent Hubble constant given by 
\begin{equation}
H(t)=\frac{\dot{R}(t)}{R(t)}
\end{equation}

\subsection{Under a dynamic discrete time}

If time is a discrete dynamical variable $\mathbf{}\mathit{\mathbf{\hat{t}}}_{n}$
the continuous space function $\mathbf{\mathrm{r}}(t)$ of a homogenous
isotropic universe is replaced by a sequence of discrete values $\mathbf{r}_{n}=\mathbf{r}\,(\mathbf{\hat{t}}_{n})$
with {[}6{]}:

\begin{equation}
(\mathbf{r}_{0},\,\mathbf{\hat{t}}_{0}),\,(\mathbf{r}_{1},\,\mathbf{\hat{t}}_{1}),........,\,(\mathbf{r}_{n},\,\mathbf{\hat{t}}_{n}),......,\,(\mathbf{r}_{N+1},\,\mathbf{\hat{t}}_{N+1})
\end{equation}
 with $(\mathbf{r}_{0},\,\mathbf{\hat{t}_{0}})$ the initial and $(\mathbf{r}_{N+1},\,\mathbf{\hat{t}}_{N+1})$
the final position. In this description $\mathbf{r}_{n}$ is still
continuous, while $\mathbf{\hat{t}}_{n}$ is discrete and as requested
a dynamic variable described by a diagonal tensor. It is noted here,
that the dynamic part of the time and its tensor character just introduced
is only needed to show the effect of the use of a continuous time
in describing the time evolution of the universe under the assumption
that time is discrete in nature with a constant step size $\Delta t\,=const$
(for example $\Delta t\,$ could be the Planck time $5.4\,10^{-44}$
s). With other words, it is assumed here that the sequence of events
happening are in general time irreversible and and can be described
by 
\[
(\mathbf{r}_{0},\,t_{0}),\,(\mathbf{r}_{1},\,t_{1}),........,\,(\mathbf{r}_{n},\,t_{n}),......,\,(\mathbf{r}_{N+1},\,t_{N+1})
\]
with $t_{n}$ being a scalar and the step size between time points
is constant with $\Delta t\,$. Thus, the exercise that follows with
a dynamic tensor description of time and the scaling variable is regarded
only a mathematical trick and not real in nature. 

The dynamic part of the time can be described by a time scaling tensor
variable denoted $\mathbf{s}{}_{n}\mathbf{1}$ defined by 

\begin{equation}
\mathbf{s}_{n}\mathbf{1}\,\Delta t\,=\mathbf{\hat{t}}_{n}-\mathbf{\hat{t}}_{n-1}
\end{equation}

with $\Delta t\,=const$. 

For simplicity reasons the system of interest is the gravitational
force of an object of mass $M$ that acts on a test particle at a
distance of radius $r$, which enables by proper selection of the
coordinate system the reduction of the dynamic part of the time to
only a single variable $s_{n}$ within the tensor notation as elaborated
on more detailed in the Appendix (please note, the Appendix also includes
the more general case). 

In discrete mechanics there are many possible definitions of the velocity
$\dot{\mathbf{r}}_{n}$. The following definition at time point $n$
follows ref {[}11{]}: 

\begin{equation}
\dot{\mathbf{r}}_{n}=\frac{\mathbf{r}_{n}-\mathbf{r}_{n-1}}{\Delta t\,}
\end{equation}

It seems to be a plausible one because it is in line with the causality
argument and our daily experience that the presence is determined
by the past and presence (i.e. to describe a present state of a system
information can experimentally-derived only from the past and presence).
With eq. (10) the velocity is defined backward in time and thus time
asymmetric, permitting a forward progressing description of the system
from past and present information. Please note, that this definition
contrasts other theories as for example quantum loop gravity theory
{[}12{]}. In presence of a discrete dynamic time variable described
by the additional time scaling variable $s_{n}$ a Lagrangian must
be derived that is able to describe adequately the system. Such a
Lagrangian has been introduced by Nose and Hoover {[}16{]}-{[}20{]}
and its discrete analog has been introduced in ref. {[}11{]}:

\begin{equation}
L_{n}^{N}=L_{n}^{N}(\mathbf{r}_{\mathrm{n}},\,\mathbf{r}_{n-1};\,\dot{s}{}_{n},\,s_{n}\mathrm{)}=s_{n}(\frac{1}{2}\,m\,\dot{\mathbf{r}}_{n}^{2}-V(\mathbf{r}_{n})+\frac{1}{2}Q\frac{\dot{s}{}_{n}^{2}}{s_{n}^{2}}-N_{df}\,k_{B}T\,ln\,s_{n})
\end{equation}

where $V(\mathbf{r}_{n})$ is the acting potential (in the case of
a Newtonian gravitational potential $V(\mathbf{r}_{n})=\text{\textgreek{F}}\,m$
), $\,k_{B}$ is the Boltzmann constant, Q is a constant with units
$Js^{2}$ (energy{*}seconds{*}seconds), which has been described as
a ``mass''-like term for the motion of $s_{n}$ with Q > 0, $N_{df}$
is the degree of freedom of the system, $\dot{s}_{n}=\frac{s_{n+1}-s_{n}}{\Delta t}$
is the velocity of the scaling factor with the unit $s^{-1}$, and
$T$ is the temperature of the system defined through the kinetic
energy of the system. A corresponding Hamiltonian without the last
two terms of eq. 11, which are not relevant here, has been described
by Elze et al. {[}10{]}.

If the discrete analog of the Lagrangian equation for each coordinate
component $i$ is defined as in ref. {[}11{]} 
\begin{equation}
\frac{\text{}1}{\Delta t}(\frac{\partial L_{n+1}^{N}(\mathbf{r}_{\mathrm{n+1}}\mathrm{,\dot{\mathbf{r}}_{n+1})}}{\mathbf{\partial\dot{\mathrm{\mathbf{\mathit{\mathrm{\mathit{r}}}}}}_{\mathrm{\mathrm{n+1}}}^{i}}}-\frac{\partial L_{n}^{N}(\mathbf{r}_{\mathrm{n}}\mathrm{,\mathrm{\dot{\mathbf{r}}_{n})}}}{\mathbf{\partial\dot{\mathrm{\mathbf{\mathit{\mathrm{\mathit{r}}}}}}_{\mathrm{\mathrm{n}}}^{i}}})=\frac{\partial L_{n}^{N}}{\partial\mathbf{\mathrm{\mathit{r}}}_{n}^{i}}
\end{equation}

the discrete Newton' law can be obtained using the discrete Lagrangian
from eq. 10

\begin{equation}
\frac{\text{}1}{\Delta t}(\dot{\,\mathbf{r}}_{n+1}\frac{s_{n+1}}{s_{n}}-\,\dot{\mathbf{r}}_{n+1}+\mathbf{\dot{r}}_{n+1}-\dot{\mathrm{\mathbf{r}}}_{n})=\frac{1}{m\,}\mathbf{F}(\mathbf{r}_{n})
\end{equation}

\begin{equation}
\ddot{\mathbf{r}}_{n}=\frac{\text{}1}{\Delta t}(\dot{\mathbf{r}}_{n+1}-\dot{\mathbf{r}}_{n})=\frac{1}{m\,}\mathbf{F}(\mathbf{r}_{n})-\dot{\mathbf{r}}_{n+1}(\frac{\mathbf{s}_{n+1}}{\mathbf{s}_{n}}-1)\frac{\text{}1}{\Delta t}
\end{equation}

\begin{equation}
\ddot{\mathbf{r}}_{n}=\frac{1}{m}\mathbf{F}(\mathbf{r}_{n})-\text{\textgreek{g}}_{n}\dot{\mathbf{r}}_{n+1}
\end{equation}

with $\text{\textgreek{g}}_{n}=\frac{\dot{s}_{n+1}}{s_{n}}$ and $\dot{s_{n}}=\frac{s_{n+1}-s_{n}}{\Delta t}$. 

If there is no scaling of time (i.e. $s_{i}=1$ for all $i=1...N+1$)
the Newton's law under a discrete time with constant time steps is
of the form $\mathbf{}m\,\ddot{\mathbf{r}}_{n}=\mathbf{F}(\mathbf{r}_{n})$ 

which resembles its continuous analog. In presence of a scaling of
time (unequal to 1) however, the Newton's law has an additional term,
which can be regarded an acceleration or a friction term in dependence
on the sign of $\text{\textgreek{g}}_{n}$. It is this friction term,
which enables discrete time physics to be time reversible with the
following request (see ref. {[}11{]}) that holds for the symmetric
case selected.

\begin{equation}
\frac{s_{n}}{s_{n+1}}=\frac{F^{i}(r_{n+1})}{F^{i}(r_{n})}\;i=x,y,z
\end{equation}
 with $\mathbf{F}(\mathbf{r}_{n})=\left[\begin{array}{c}
F^{x}(\mathbf{r}_{n})\\
F^{y}(\mathbf{r}_{n})\\
F^{z}(\mathbf{r}_{n})
\end{array}\right]=\left[\begin{array}{c}
F^{x}(\mathbf{r}_{n})\\
F^{x}(\mathbf{r}_{n})\\
F^{x}(\mathbf{r}_{n})
\end{array}\right]$.

This equation (eq. 16) and its more general analog in the Supplementary
Material (eq. 54) are called the reversibility axiom. If the reversibility
axiom is fulfilled, the introduced discrete time physics is time reversible
{[}11{]}.

By incorporating the reversibility axiom into the discrete Newtonian
equation (eq. 15) the following expression is obtained:

\begin{equation}
\ddot{\mathbf{r}}_{n}=\frac{\text{}1}{\Delta t}(\dot{\mathbf{r}}_{n+1}-\dot{\mathbf{r}}_{n})=\frac{1}{m\,}\mathbf{F}(\mathbf{r}_{n})+\dot{\mathbf{r}}_{n+1}\frac{\dot{F^{x}}(r_{n+1})}{F^{x}(r_{n+1})}
\end{equation}

with $\dot{F^{x}}(r_{n+1})=\frac{F^{x}(r_{n+1})-F^{x}(r_{n})}{\Delta t}$
. Eq. 17 reflects thereby the time reversible evolution of a system
under a discrete time. It is highlighted that due to the dynamic nature
of the discrete time with the variable $s_{n}$ time reversibility
is guaranteed. 

As in the case of Newton's law (eq. 17) the introduction of a dynamic
discrete time has an impact also on other time-dependent formulas
such as the well known differential equation $\frac{\dot{x}}{x}=b$
with $b=const$ having as solution the exponential function $x(t)=x_{0}\,e^{b\,t}$
with $x_{0}$ the value at time 0. The corresponding differential
equation under a discrete dynamic time (dependent on eq. 16) is given
by 

\begin{equation}
\frac{\dot{x}_{n}}{x_{n}}=b\,s_{n}
\end{equation}

This can be derived from $x_{n}(t)=x_{0}\prod_{i=1}^{n}(1+bs_{i}\Delta t)\approx x_{0}\,e^{b\,{\displaystyle \sum_{i=1}^{n}s_{i}\Delta t}}$
with the discrete definition of $\dot{x}_{n}$ given above for $\dot{\mathbf{r}}_{n}$
(eq. 10). The latter equation is true at the limit $lim\,\Delta t\rightarrow0$. 

\subsection{Under a dynamic continuous time }

In order to elaborate on the consequences of using a time continuous
description of the evolution of the universe if time is actually discrete
and non dynamic in nature, the above description with a dynamic discrete
time is set to the limit with $lim\:\Delta t\rightarrow0$. 

For this, the derived discrete Newtonian equation (eq. 17) is transformed
into its corresponding continuous analog:

\begin{equation}
\ddot{\mathbf{r}}=\frac{1}{m}\mathbf{F}(\mathbf{r})-\text{\textgreek{g}}\dot{\mathbf{r}}
\end{equation}

with $\text{\textgreek{g}}=\frac{\dot{s}}{s}=-\frac{\dot{F^{x}}}{F^{x}}$
with $s(t)$. 

Correspondingly, in this time continuous frame the differential equation
of the exponential function is then given by $\frac{\dot{x}}{x}=b\,s(t)$
and thus the Hubble law of eq. 5 is given by

\begin{equation}
\frac{\dot{R}}{R}=H(t)\,s(t)
\end{equation}

Please note, within the system of interest, which is homogenous isotropic
expanding universe with the argumentation put forward in eq. 22 below,
the $s$in eq. 19 and eq. 20 are the same. The introduction of the
dynamic continuous time resolves also the problem of the immediate
request for a discrete space that comes along with a discrete time
through the Lorentz transformation as has been done for example by
the quantum loop gravity theory {[}12{]}. However, by the present
translation from the discrete time to a dynamic continuous time with
$lim\:\Delta t\rightarrow0$ the space can be described as usual because
no need for a dynamic continuous space is required as space has not
the problem of irreversibility.

\subsection{From the Poisson equation to the Evolution of the Universe }

Applying the continuous formulation of the modified Newton's equation
of eq. 17 to the gravitational force with $\mathbf{F}(\mathbf{r})=\mathbf{F}_{G}(\mathbf{r})$
for the system of interest here (using $\dot{F_{G}^{i}}(r)=-2\frac{G\,M\,m}{r^{i3}}\,\dot{r^{i}}$)
the following equation is obtained:

\begin{equation}
\ddot{\mathbf{r}}=\frac{1}{m\,}\mathbf{F}(\mathbf{r})-2\frac{\dot{r_{x}}}{r_{x}}\dot{\mathbf{r}}=-\frac{GM}{r^{2}\,}\frac{\mathbf{r}}{|\mathbf{r}|}+2(\frac{\dot{r_{x}}}{r_{x}})^{2}\,\mathbf{r}
\end{equation}

In an isotropic homogenous and spherically expanding universe at the
Newtonian limit (eq. 5: $\ \ \,\ddot{\mathbf{r}}=-\frac{G\,M\,}{\mathbf{r}^{2}}\:\frac{\mathbf{r}}{|\mathbf{r}|}+\frac{1}{3}\text{\textgreek{L}}\,\mathbf{r}$)
the scaling factor $R$ is proportional to $r$and $\dot{R}$ proportional
to $\dot{r}$ and thus we obtain for the second term of eq. 19 (using
also eq. 20) 

\begin{equation}
2\,(\frac{\dot{r_{x}}}{r_{x}})^{2}=2\,(\frac{\dot{R}}{R})^{2}=2\,H(t)^{2}\,s(t)^{2}\,
\end{equation}
which in combination with the Newtonian limit of the poisson equation
derived for the general relativity theory (eq. 3) yields 

\begin{equation}
\frac{1}{3}\text{\textgreek{L}}=2\,H(t)^{2}s(t)^{2}\,
\end{equation}
This results under a slow force changing limit (i.e. $s\approx1$
) to

\begin{equation}
\text{\textgreek{L}}=6\,H(t)^{2}
\end{equation}

or under a constant changing force with $s(t)=\sqrt{1/2}$ 

\begin{equation}
H=\text{\ensuremath{\sqrt{\frac{\text{\textgreek{L}}}{3}}}}
\end{equation}

usually obtained under standard descriptions {[}14{]} (obviously $s(t)=\sqrt{1/2}$
was chosen by purpose albeit meaningless and not required for the
further derivations; alternatively, the geometry of the expanding
universe can be altered such that $2\,(\frac{\dot{r_{x}}}{r_{x}})^{2}=\,(\frac{\dot{R}}{R})^{2}$). 

Concluding the first part of the presented theory, when time continuous
physics at the Newtonian limit is applied to a flat isotropic homogenous
universe that evolves in discrete time steps under slow changing force,
it appears that the universe is expanding with a positive cosmic constant
of \textgreek{L} = $6\,H^{2}$. With other words, if time is discrete,
the cosmic constant is a consequence of applying time reversible continuous
physics to the expansion of the universe. It is further interesting
to note, that under the given condition with the cosmic constant being
constant by definition, the Hubble constant is independent of time
if the gravitational force does not and never did change fast.

\subsection{The expansion of the universe with its apparent inflation}

The condition that the gravitational force is not changing fast is
of course not correct in the early time of the expansion of the universe,
which is studied next. For this purpose we use the following relationship
$\text{\textgreek{g}}=\frac{\dot{s}}{s}=-\frac{\dot{F}}{F}=\frac{\dot{r}}{r}=\frac{\dot{R}(t)}{R(t)}$
yielding

\begin{equation}
\frac{\dot{s}(t)}{s(t)}=\frac{\dot{R}(t)}{R(t)}=H(t)\,s(t)=\sqrt{\frac{\text{\textgreek{L}}}{6}}
\end{equation}

and correspondingly 
\begin{equation}
s(t)=s_{0}\,e^{\sqrt{\frac{\text{\textgreek{L}}}{6}}\,t}
\end{equation}

with $s_{0}$ the scaling at time 0 (please note, that this $s_{0}$
is from the continuous description of time, while in the discrete
description $s_{n}$ i used with $n$ an integer starting from 1)

Since 
\begin{equation}
H(t)=\frac{\sqrt{\frac{\text{\textgreek{L}}}{6}}}{s(t)}=\sqrt{\frac{\text{\textgreek{L}}}{6}}\frac{1}{s_{0}}e^{-\sqrt{\frac{\text{\textgreek{L}}}{6}}\,t}
\end{equation}
If one now takes the Hubble constant of eq. 36 and applies it for
the determination of the scaling factor by using the standard Hubble
law of eq. 5 we obtain the following scaling factor of the metric
of the universe
\begin{equation}
R(t)=R_{0}e^{\int_{0}^{t}\frac{\sqrt{\frac{\text{\textgreek{L}}}{6}}}{s(t)}dt}=R_{0}e^{\frac{\sqrt{\frac{\text{\textgreek{L}}}{6}}}{s_{0}}\int_{0}^{t}e^{-\sqrt{\frac{\text{\textgreek{L}}}{6}}\,t}dt}=R_{0}e^{\frac{1}{s_{0}}[1-e^{-\sqrt{\frac{\text{\textgreek{L}}}{6}}\,t}]}
\end{equation}

Hence, when time continuous physics is applied to a flat isotropic
homogenous universe that evolves in discrete time steps there is an
apparent expansion of the universe following eq. 29. This artifact
is constant for large time $t$$lim_{t\rightarrow\infty}R(t)=R_{0}\,e^{\frac{1}{s_{0}}}$
, while for small times using a Taylor expansion 1. order of the exponential
function the universe appears to expand exponentially with $R(t)\approx R_{0}\,e^{\frac{\sqrt{\frac{\text{\textgreek{L}}}{6}}}{\,s_{0}}\,t}$
as postulated by Guth {[}14{]}.

There are three so far three undetermined constants left in eq. 29
(i.e. $R_{0},\:s_{0}$ , and $\text{\textgreek{L}}$ ) that need to
be discussed for a quantitative time continuous description of the
evolution of the universe taking into account that time is actually
discrete and irreversible. First, $R_{0}$ is considered. Because
the scaling factor $R(t)$ is in frame of the reference at time point
0, $R_{0}=1$. Next, the unknown constant $s_{0}$ is determined by
assuming that the first time step of the universe is the Planck time
$\Delta t_{p}=\sqrt{\frac{hG}{c^{5}}=}5.3\,10^{-44}s$ (with $h$
Planck's Wirkungsquantum and $c$the velocity of light in vacuum)
and that the universe was expanding with maximal velocity, i.e. light
velocity $c$. Furthermore, it is assumed that the universe at time
point $t=0$ had the minimal size of a system possible, which is regarded
the Planck length $l_{p}.$ Since by definition $l_{p}=c\,\Delta t_{p}$
at time point $t_{1}$ the size of the universe was thus $l_{p}+c\,\Delta t_{p}=2\,l_{p}=R(t_{1})\,l_{p}$
yielding $R(t_{1})=2$. We further describe $R(t_{1})$ by a Taylor
expansion first order and using $\dot{R}(t)=R(t)\,e^{\sqrt{\frac{\text{\textgreek{L}}}{6}}t}\,\sqrt{\frac{\text{\textgreek{L}}}{6}}\,\frac{1}{s_{0}}$
starting with time $t=0$ yielding 

\begin{equation}
R(t_{1})\approx R_{0}+\dot{R}(0)\Delta t_{p}=1+\sqrt{\frac{\text{\textgreek{L}}}{6}}\,\frac{1}{s_{0}}\Delta t_{p}
\end{equation}
which results in 

\begin{equation}
s_{0}=\sqrt{\frac{\text{\textgreek{L}}}{6}}\,\Delta t_{p}
\end{equation}

since $R(t_{1})=2$. Following these assumptions, the quantitative
expression of the expansion of the universe is given by
\begin{equation}
R(t)=e^{\frac{1}{\sqrt{\frac{\text{\textgreek{L}}}{6}}\,\Delta t_{p}}[1-e^{-\sqrt{\frac{\text{\textgreek{L}}}{6}}\,t}]}
\end{equation}

The scaling factor for large time $t$is thus given by $lim_{t\rightarrow\infty\,}R(t)=R_{inf}=\,e^{\frac{\sqrt{\frac{6}{\Lambda}}}{\Delta t_{p}}}$
and constant, while for small times using a Taylor expansion 1. order
of the exponential function the universe appears to expand exponentially
with $R(t)\approx\,e^{\frac{t}{\Delta t_{p}}\,}$ as requested by
Guth {[}14{]}. The last unknown is the cosmic constant $\text{\textgreek{L}}$,
which can be gathered from measurements and the Hubble law with an
$\text{\textgreek{L}}$ in the order of $10^{-35}s^{-2}$ assumed
in the following to be $3*10^{-35}s^{-2}${[}15{]}. This allows to
calculate the evolution of the universe using eq. 32 as demonstrated
in Figure 1. Figure 1 shows that when time continuous physics is applied
to a flat isotropic homogenous universe that evolves in discrete time
steps, it appears that the universe is expanding through an inflationary
phase during the first $3\:10^{-35}s$ after which it approximates
its maximal expansion at an R(t) of ca $10^{60}=R_{inf}$ . Please
note, that the presented approach does neither take special relativity
nor quantum effects into account, which in particular at early time
points such as during the grand unification theory (GUT) time, may
have to be considered. Nevertheless, assuming a universe size of $1\,l_{p}$
at time point 0 the expansion of the universe follows closely accepted
inflationary universe models {[}14{]},{[}15{]}. Furthermore, the model
predicts well the current size of the universe having $R(t_{present})\approx R_{inf}=4\,10^{60}$
yielding a size of the universe of $R_{inf}\,l_{p}=R_{inf}\,c\,\Delta t_{p}\approx1\,10^{27}m$.
Of course, the return process starting on the measured current size
of the universe in combination with eq. 32 is also possible and results
in the cosmic constant with expected size. 

\includegraphics[scale=0.8]{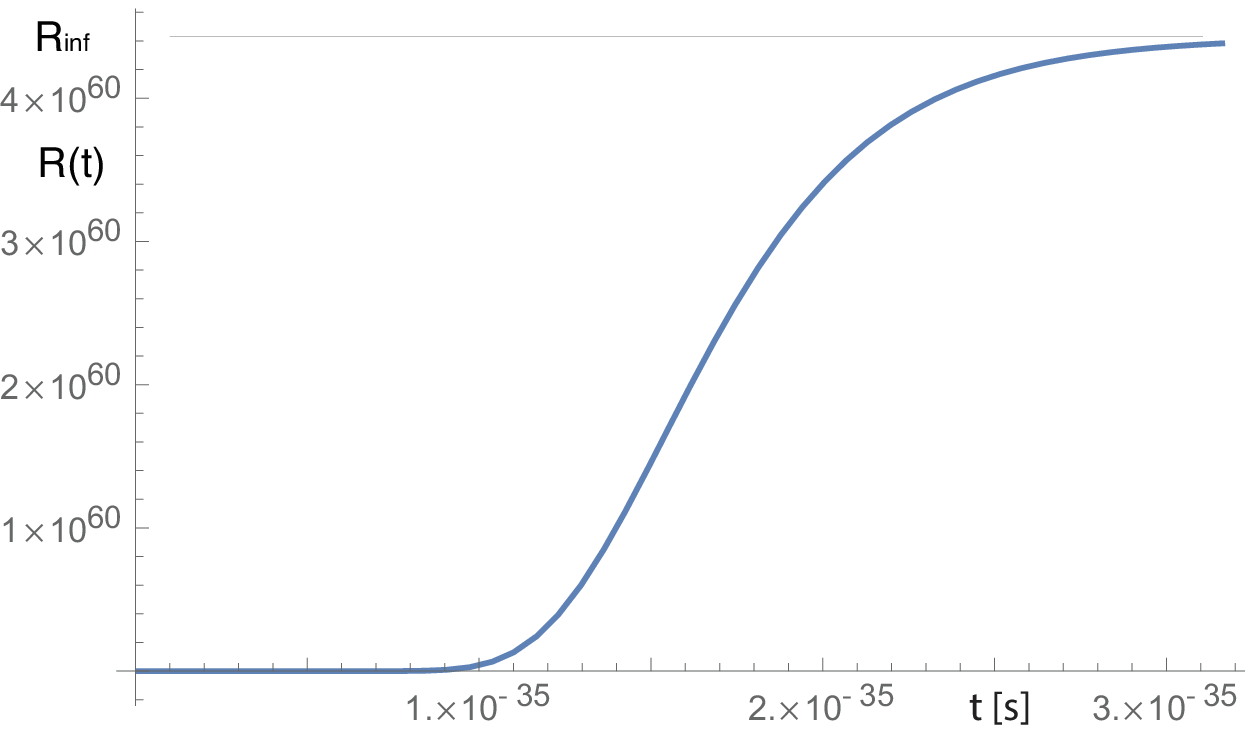}

Figure 1: The scaling factor $R(t)$ versus time $t$using eq. 32.
In addition, the maximal expansion factor $lim_{t\rightarrow\infty}R(t)=\,e^{\frac{\sqrt{\frac{6}{\text{\textgreek{L}}}}}{\Delta t_{p}}}$
is indicated.

\subsection{The apparent acceleration of the universe at late time points}

Using the standard model of general relativity in combination with
experimental measures it is often suggested that the universe is currently
in an accelerating phase {[}15{]}. The presented description of the
universe assuming a continuous time while time is postulated discrete
in nature yields however a maximal scale factor of $lim_{t\rightarrow\infty}R(t)=\,e^{\frac{\sqrt{\frac{6}{\text{\textgreek{L}}}}}{\Delta t_{p}}}$
(see also Figure 1). While it depends on the cosmic constant $\text{\textgreek{L}}$,
within the description presented the size of the cosmic constant does
not determine whether the universe is accelerating in size, expanding
forever, or may collapse. Within this model, there is thus no support
and no requirement for the proclaimed current accelerated universe. 

\section{Conclusion}

The present work studied the artifacts that appear when time continuous
reversible physics is applied to a universe that evolves irreversibly
under a discrete time. The artifacts are (i) a cosmic constant and
(ii) an inflationary evolution of the universe at early time - two
prominent postulates in cosmology of which physical origins and requests
are puzzling. Several recent attempts to resolve these issues have
been suggested as for example {[}22-24{]}. The presented work suggests
a simple solution by demanding a discrete time. The consequences are
manyfold such as that the present universe is not expanding under
acceleration as believed, nor that an inflationary expansion of the
universe is necessary to describe the evolution of the universe. We
invite the reader to apply the concept of a discrete time to further
areas in physics and cosmology.

\section{Acknowledgment}

We would like to thank the ETH for unrestrained financial support
and Dr. Yaron Harad, Dr. Alexander Sobol, Dr. Witek Kwiatkowski, and
Dr. Juerg Froehlich for helpful discussions.

\section{Appendix}

The following time reversible discrete time physics has been published
in ref {[}11{]} and is repeated here for the example of interest. 

If time is a discrete dynamical variable $\mathbf{}\mathit{\mathbf{\hat{t}}}_{n}$
the continuous space function $\mathbf{\mathrm{r}}(t)$ of a homogenous
isotropic universe is replaced by a sequence of discrete values $\mathbf{r}_{n}=\mathbf{r}\,(\mathbf{\hat{t}}_{n})$
with {[}6{]}:

\begin{equation}
(\mathbf{r}_{0},\,\mathbf{\hat{t}}_{0}),\,(\mathbf{r}_{1},\,\mathbf{\hat{t}}_{1}),........,\,(\mathbf{r}_{n},\,\mathbf{\hat{t}}_{n}),......,\,(\mathbf{r}_{N+1},\,\mathbf{\hat{t}}_{N+1})
\end{equation}
 with $(\mathbf{r}_{0},\,\mathbf{\hat{t}_{0}})$ the initial and $(\mathbf{r}_{N+1},\,\mathbf{\hat{t}}_{N+1})$
the final position. In this description $\mathbf{r}_{n}$ is still
continuous, while $\mathbf{\hat{t}}_{n}$ is discrete and as requested
a dynamic variable described by a diagonal tensor second order. It
is noted here, that the dynamic part of the time and its tensor character
just introduced is only needed to show the effect of the use of a
continuous time in describing the time evolution of the universe under
the assumption that time is discrete in nature with a constant step
size $\Delta t\,=const$ (for example $\Delta t\,$ could be the Planck
time $5.4\,10^{-44}$ s). With other words, it is assumed here that
the sequence of events happening are in general time irreversible
and and can be described by 
\[
(\mathbf{r}_{0},\,t_{0}),\,(\mathbf{r}_{1},\,t_{1}),........,\,(\mathbf{r}_{n},\,t_{n}),......,\,(\mathbf{r}_{N+1},\,t_{N+1})
\]
with $t_{n}$ being a scalar and the step size between time points
is constant with $\Delta t\,$. Thus, the exercise that follows with
a dynamic tensor description of time and the scaling variable is regarded
only a mathematical trick and not real in nature. 

The dynamic part of the time can be described by a time scaling tensor
variable denoted $\mathbf{s}{}_{n}\mathbf{1}$ defined by 

\begin{equation}
\mathbf{s}_{n}\mathbf{1}\,\Delta t\,=\mathbf{\hat{t}}_{n}-\mathbf{\hat{t}}_{n-1}
\end{equation}

with $\Delta t\,=const$. In matrix notation this reads:

\begin{equation}
\left[\begin{array}{c}
s_{n}^{x}\\
s_{n}^{y}\\
s_{n}^{z}
\end{array}\right]\left[\begin{array}{ccc}
1 & 0 & 0\\
0 & 1 & 0\\
0 & 0 & 1
\end{array}\right]\Delta t\,=\left[\begin{array}{ccc}
t_{n}^{x}-t_{n-1}^{x} & 0 & 0\\
0 & t_{n}^{y}-t_{n-1}^{y} & 0\\
0 & 0 & t_{n}^{z}-t_{n-1}^{z}
\end{array}\right]=\left[\begin{array}{ccc}
t_{n}^{x} & 0 & 0\\
0 & t_{n}^{y} & 0\\
0 & 0 & t_{n}^{z}
\end{array}\right]-\left[\begin{array}{ccc}
t_{n-1}^{x} & 0 & 0\\
0 & t_{n-1}^{y} & 0\\
0 & 0 & t_{n-1}^{z}
\end{array}\right]
\end{equation}

For simplicity reasons the system of interest is the gravitational
force of an object of mass $M$ that acts on a test particle at a
distance of radius $r$, which enables by proper selection of the
coordinate system to reduce the dynamic part of the time to only a
single variable $s_{n}$ within the tensor notation. 

\begin{equation}
\left[\begin{array}{c}
s_{n}^{x}\\
s_{n}^{y}\\
s_{n}^{z}
\end{array}\right]\left[\begin{array}{ccc}
1 & 0 & 0\\
0 & 1 & 0\\
0 & 0 & 1
\end{array}\right]\Delta t\,=s_{n}1\Delta t
\end{equation}

The more general case is treated below. 

In discrete mechanics there are many possible definitions of the velocity
$\dot{\mathbf{r}}_{n}$. The following definition at time point $n$
seems to be a plausible definition: 

\begin{equation}
\dot{\mathbf{r}}_{n}=\frac{\mathbf{r}_{n}-\mathbf{r}_{n-1}}{\Delta t\,}
\end{equation}

because it is in line with the causality argument and our daily experience
that the presence is determined by the past and presence (i.e. to
describe a present state of a system information can experimentally-derived
only from the past and presence). With eq. (37) the velocity is defined
backward in time and thus time asymmetric, permitting a forward progressing
description of the system from past and present information. Please
note, that this definition contrasts other theories as for example
quantum loop gravity theory {[}12{]}. In presence of a discrete dynamic
time variable described by the additional time scaling variable $s_{n}$
a Lagrangian must be derived that is able to describe adequately the
system. Such a Lagrangian under an artificial continuous scaled time
$s$ has been introduced by Nose and Hoover ($L^{N}=s(\frac{1}{2}\,m\,\dot{r}^{2}-V(r)+\frac{1}{2}Q\frac{\dot{s}{}^{2}}{s^{2}}-N_{df}\,k_{B}T\,ln\,s)$){[}16{]}-{[}20{]}.
The discrete analog of the Nose-Hoover Lagrangian (in real space)
is given by the following expression {[}11{]}:

\begin{equation}
L_{n}^{N}=L_{n}^{N}(\mathbf{r}_{\mathrm{n}},\,\mathbf{r}_{n-1};\,\dot{s}{}_{n},\,s_{n}\mathrm{)}=s_{n}(\frac{1}{2}\,m\,\dot{\mathbf{r}}_{n}^{2}-V(\mathbf{r}_{n})+\frac{1}{2}Q\frac{\dot{s}{}_{n}^{2}}{s_{n}^{2}}-N_{df}\,k_{B}T\,ln\,s_{n})
\end{equation}

where $V(\mathbf{r}_{n})$ is the acting potential (in the case of
a Newtonian gravitational potential $V(\mathbf{r}_{n})=\text{\textgreek{F}}\,m$
), $\,k_{B}$ is the Boltzmann constant, Q is a constant with units
$Js^{2}$ (energy{*}seconds{*}seconds), which has been described as
a ``mass''-like term for the motion of $s_{n}$ with Q > 0, $N_{df}$
is the degree of freedom of the system, $\dot{s}_{n}=\frac{s_{n+1}-s_{n}}{\Delta t}$
is the velocity of the scaling factor with the unit $s^{-1}$, and
$T$ is the temperature of the system defined through the kinetic
energy of the system. A corresponding Hamiltonian without the last
two terms of eq. 38 has been described by Elze et al. {[}10{]}. 

If the discrete analog of the Lagrangian equation for each coordinate
component $i$ is defined as in ref. {[}11{]} 
\begin{equation}
\frac{\text{}1}{\Delta t}(\frac{\partial L_{n+1}^{N}(\mathbf{r}_{\mathrm{n+1}}\mathrm{,\dot{\mathbf{r}}_{n+1})}}{\mathbf{\partial\dot{\mathrm{\mathbf{\mathit{\mathrm{\mathit{r}}}}}}_{\mathrm{\mathrm{n+1}}}^{i}}}-\frac{\partial L_{n}^{N}(\mathbf{r}_{\mathrm{n}}\mathrm{,\mathrm{\dot{\mathbf{r}}_{n})}}}{\mathbf{\partial\dot{\mathrm{\mathbf{\mathit{\mathrm{\mathit{r}}}}}}_{\mathrm{\mathrm{n}}}^{i}}})=\frac{\partial L_{n}^{N}}{\partial\mathbf{\mathrm{\mathit{r}}}_{n}^{i}}
\end{equation}

the discrete Newton' law can be obtained using the discrete Lagrangian
from eq. 10

\begin{equation}
\frac{\text{}1}{\Delta t}(\dot{\,\mathbf{r}}_{n+1}\frac{s_{n+1}}{s_{n}}-\,\dot{\mathbf{r}}_{n+1}+\mathbf{\dot{r}}_{n+1}-\dot{\mathrm{\mathbf{r}}}_{n})=\frac{1}{m\,}\mathbf{F}(\mathbf{r}_{n})
\end{equation}

\begin{equation}
\ddot{\mathbf{r}}_{n}=\frac{\text{}1}{\Delta t}(\dot{\mathbf{r}}_{n+1}-\dot{\mathbf{r}}_{n})=\frac{1}{m\,}\mathbf{F}(\mathbf{r}_{n})-\dot{\mathbf{r}}_{n+1}(\frac{\mathbf{s}_{n+1}}{\mathbf{s}_{n}}-1)\frac{\text{}1}{\Delta t}
\end{equation}

\begin{equation}
\ddot{\mathbf{r}}_{n}=\frac{1}{m}\mathbf{F}(\mathbf{r}_{n})-\text{\textgreek{g}}_{n}\dot{\mathbf{r}}_{n+1}
\end{equation}

with $\text{\textgreek{g}}_{n}=\frac{\dot{s}_{n+1}}{s_{n}}$ and $\dot{s_{n}}=\frac{s_{n+1}-s_{n}}{\Delta t}$. 

An alternative derivation results in 

\begin{equation}
\ddot{\mathbf{r}}_{n}=\frac{s_{n}}{m\,s_{n+1}}\mathbf{F}(\mathbf{r}_{n})-\dot{\mathbf{r}}_{n}(1-\frac{s_{n}}{s_{n+1}})\frac{1}{\Delta t}
\end{equation}

If there is no scaling of time (i.e. $s_{i}=1$ for all $i=1...N+1$)
the Newton's law under a discrete time with constant time steps is
of the form $\mathbf{}m\,\ddot{\mathbf{r}}_{n}=\mathbf{F}(\mathbf{r}_{n})$ 

which resembles its continuous analog. In presence of a scaling of
time unequal to 1 however, the Newton's law has an additional term,
which can be regarded an acceleration or a friction term in dependence
on the sign of $\text{\textgreek{g}}_{n}$. It is this friction term,
which enables discrete time physics to be time reversible as demonstrated
in the following.

Time reversibility can be described by a two step process having one
step forward followed by a step backward. Let us consider the evolution
of the discrete Newton's law with two steps forwards. Following eq.
14 
\global\long\def\labelenumi{(\roman{enumi})}%
 by solving $\ddot{\mathbf{r}}_{n}=\frac{\text{}1}{\Delta t}(\dot{\mathbf{r}}_{n+1}-\dot{\mathbf{r}}_{n})$
for 
\begin{enumerate}
\item 
\begin{equation}
\dot{\mathbf{r}}_{n+1}=\dot{\mathbf{r}}_{n}+\frac{s_{n}}{m\,s_{n+1}}\Delta t\mathbf{\,F}(r_{n})-(1-\frac{s_{n}}{s_{n+1}})\dot{\mathrm{r}}_{n}
\end{equation}
 For the second time step we take the second expression of the discrete
Newton's law (eq. 13)
\item 
\begin{equation}
\dot{\mathbf{r}}_{n+2}=\dot{\mathbf{r}}_{n+1}+\frac{1}{m}\Delta t\,\mathbf{F}(\mathbf{r}_{n+1})-\dot{\mathbf{r}}_{n+2}(\frac{s_{n+2}}{s_{n+1}}-1)
\end{equation}
\end{enumerate}
If the second step is now backward in time
\begin{equation}
\dot{\mathbf{r}}_{n+2}=\dot{\mathbf{r}}_{n+1}-\frac{1}{m}\Delta t\,\mathbf{F}(\mathbf{r}_{n+1})-\dot{\mathbf{r}}_{n+2}(\frac{s_{n+2}}{s_{n+1}}-1)
\end{equation}

and if time reversibility is requested (i.e. $\dot{\mathbf{r}}_{n}=\dot{\mathbf{r}}_{n+2}$
and $s_{n+2}=s_{n}$)

\begin{equation}
\mathbf{\dot{r}}_{n}=\dot{\mathbf{r}}_{n+2}=\dot{\mathbf{r}}_{n+1}-\frac{1}{m}\Delta t\,\mathbf{F}(\mathbf{r}_{n+1})-\dot{\mathbf{r}}_{n}(\frac{s_{n}}{s_{n+1}}-1)
\end{equation}

\begin{equation}
\dot{\mathbf{r}}_{n}=\dot{\mathbf{r}}_{n}+\frac{s_{n}}{m\,s_{n+1}}\Delta t\,\mathbf{F}(\mathbf{r}_{n})-(1-\frac{s_{n}}{s_{n+1}})\dot{\mathbf{r}}_{n}-\frac{1}{m}\Delta t\,\mathbf{F}(\mathbf{r}_{n+1})-\dot{\mathbf{r}}_{n}(\frac{s_{n}}{s_{n+1}}-1)
\end{equation}

\begin{equation}
\frac{s_{n}}{s_{n+1}}\mathbf{F}(\mathbf{r}_{n})=\mathbf{F}(\mathbf{r}_{n+1})
\end{equation}

and thus

\begin{equation}
\frac{s_{n}}{s_{n+1}}=\frac{F^{i}(r_{n+1})}{F^{i}(r_{n})}\;i=x,y,z
\end{equation}
 with $\mathbf{F}(\mathbf{r}_{n})=\left[\begin{array}{c}
F^{x}(\mathbf{r}_{n})\\
F^{y}(\mathbf{r}_{n})\\
F^{z}(\mathbf{r}_{n})
\end{array}\right]=\left[\begin{array}{c}
F^{x}(\mathbf{r}_{n})\\
F^{x}(\mathbf{r}_{n})\\
F^{x}(\mathbf{r}_{n})
\end{array}\right]$.

This equation holds for the symmetric case selected. 

This equation (eq. 50) and its more general analog below (eq. 55)
are called the reversibility axiom. If the reversibility axiom is
fulfilled, the introduced discrete time physics is time reversible.

By incorporating the reversibility axiom on the symmetric case selected
(eq. 50) into the discrete Newtonian equation (eq. 42) the following
expression is obtained:

\begin{equation}
\ddot{\mathbf{r}}_{n}=\frac{\text{}1}{\Delta t}(\dot{\mathbf{r}}_{n+1}-\dot{\mathbf{r}}_{n})=\frac{1}{m\,}\mathbf{F}(\mathbf{r}_{n})+\dot{\mathbf{r}}_{n+1}\frac{\dot{F^{x}}(r_{n+1})}{F^{x}(r_{n+1})}
\end{equation}

with $\dot{F^{x}}(r_{n+1})=\frac{F^{x}(r_{n+1})-F^{x}(r_{n})}{\Delta t}$
. Eq. 51 reflects thereby the time reversible evolution of a system
under a discrete time. It is thereby highlighted that due to the dynamic
nature of the discrete time with the variable $s_{n}$ time reversibility
was obtained. 

The approach taken above has been focusing on a spherical potential
that acts along all the coordinates the same for simplicity. Here,
the more general case is highlighted of an acting force on a single
particle starting with time to be a discrete dynamical tensor second
order $\mathbf{}\mathit{\mathbf{\hat{t}}}_{n}$ describing a series
of events as follows

\begin{equation}
(\mathbf{r}_{0},\,\mathbf{\hat{t}}_{0}),\,(\mathbf{r}_{1},\,\mathbf{\hat{t}}_{1}),........,\,(\mathbf{r}_{n},\,\mathbf{\hat{t}}_{n}),......,\,(\mathbf{r}_{N+1},\,\mathbf{\hat{t}}_{N+1})
\end{equation}
 with $(\mathbf{r}_{0},\,\mathbf{\hat{t}_{0}})$ the initial and $(\mathbf{r}_{N+1},\,\mathbf{\hat{t}}_{N+1})$
the final position. the dynamical part of time can then be described
by the scaling factor 
\begin{equation}
\mathbf{\hat{s}}_{n}\Delta t=\mathbf{s}_{n}\mathbf{1}\,\Delta t\,=\left[\begin{array}{ccc}
s_{n}^{x} & 0 & 0\\
0 & s_{n}^{y} & 0\\
0 & 0 & s_{n}^{z}
\end{array}\right]\Delta t\,=\left[\begin{array}{ccc}
t_{n}^{x} & 0 & 0\\
0 & t_{n}^{y} & 0\\
0 & 0 & t_{n}^{z}
\end{array}\right]-\left[\begin{array}{ccc}
t_{n-1}^{x} & 0 & 0\\
0 & t_{n-1}^{y} & 0\\
0 & 0 & t_{n-1}^{z}
\end{array}\right]=\mathbf{\hat{t}}_{n}-\mathbf{\hat{t}}_{n-1}
\end{equation}

This yields the following reversibility axiom

\begin{equation}
\frac{s_{n}^{x}}{s_{n+1}^{x}}=\frac{F^{x}(r_{n+1})}{F^{x}(r_{n})},\:\frac{s_{n}^{y}}{s_{n+1}^{y}}=\frac{F^{y}(r_{n+1})}{F^{y}(r_{n})},\:\frac{s_{n}^{z}}{s_{n+1}^{z}}=\frac{F^{z}(r_{n+1})}{F^{z}(r_{n})}
\end{equation}

By incorporating the reversibility axiom into the discrete Newtonian
equation (eq. 42) the following expression is obtained:

\begin{equation}
\ddot{\mathbf{r}}_{n}=\frac{\text{}1}{\Delta t}(\dot{\mathbf{r}}_{n+1}-\dot{\mathbf{r}}_{n})=\frac{1}{m\,}\mathbf{F}(r_{n})-\dot{\mathbf{r}}_{n+1}\mathbf{}\left[\begin{array}{ccc}
(\frac{F^{x}(r_{n})}{F^{x}(r_{n+1})}-1)\frac{1}{\Delta t} & 0 & 0\\
0 & (\frac{F^{y}(r_{n})}{F^{y}(r_{n+1})}-1)\frac{1}{\Delta t} & 0\\
0 & 0 & (\frac{F^{z}(r_{n})}{F^{z}(r_{n+1})}-1)\frac{1}{\Delta t}
\end{array}\right]
\end{equation}

Next, the derived discrete Newtonian equation is transformed into
its corresponding continuous analog by $lim\,\Delta t\rightarrow0$:

\begin{equation}
\ddot{\mathbf{r}}=\frac{1}{m}\mathbf{F}(\mathbf{r})-\text{\ensuremath{\hat{\mathbf{\mathbf{\gamma}}}}}\dot{\mathbf{\,r}}
\end{equation}

with 

\[
\hat{\mathbf{\mathbf{\gamma}}}=\left[\begin{array}{ccc}
-\frac{\dot{F^{x}}}{F^{x}} & 0 & 0\\
0 & -\frac{\dot{F^{y}}}{F^{y}} & 0\\
0 & 0 & -\frac{\dot{F^{z}}}{F^{z}}
\end{array}\right]
\]

\section{References}

{[}1{]} Farias RAH and Recami E 2007 arXiv:quant-ph/9706059

{[}2{]} Thomson JJ 1925 \textit{Proc. Roy. Soc. of Edinburgh} 46 \textbf{90}

{[}3{]} Yang CN 1947 \textit{Phys. Rev.} 72 \textbf{874}

{[}4{]} Levi R 1927\textit{ Journal de Physique et le Radium} 8, \textbf{182}

{[}5{]} Caldirola P 1953 \textit{Supplmento al Nuovo Cimento} 10,
\textbf{1747}

{[}6{]} Lee TD 1983 \textit{Physics Lett.} 122B \textbf{217}

{[}7{]} 't Hooft G 2010 \textit{Int. J. Mod. Phys.} A 25 \textbf{4385}

{[}8{]} 't Hooft G 2014 arXiv:1405.1548

{[}9{]} Elze HT 2013 arXiv:1310.2862

{[}10{]} Elze HT 2014 \textit{Phys. Rev. A} 89 \textbf{012111}.

{[}11{]} Riek R 2014 \textit{Entropy},16 \textbf{3149}.

{[}12{]} Rovelli C 2011 arXiv:11-2.3660

{[}13{]} Einstein A 1917 \textit{Sitzungsberichte der Koniglich Preussischen
Akademie der Wissenschaften Berlin}. part 1: \textbf{142}.

{[}14{]} Guth A 1981 \textit{Phys. Rev. D }23\textbf{ 347}

{[}15{]} Straumann N 2013 \textit{General Relativity}, Springer, Dordrecht

{[}16{]} Nowakowski M 2001 \textit{Int. J. Mod. Phys. D. }10 \textbf{649}

{[}17{]} Nose S 1984 \textit{J. Chem. Phys.} 81 \textbf{511}

{[}18{]} Nose S 1986\textit{ Mol. Phys. 57} \textbf{187}

{[}19{]} Hoover WG 1985\textit{ Phys. Rev. A }31 \textbf{1695}

{[}20{]} Hunenberger PH 2005 \textit{Adv. Polym. Sci.} 173 \textbf{105}

{[}21{]} Hubble E 1929 \textit{Proc. Natl. Acad. Sci.} 15 \textbf{168} 

{[}22{]} Verlinde E 2016 arXiv:1611.02269v2

{[}23{]} Hu W and Sawicki I 2007 \textit{Phys. Rev. D} 76 \textbf{064004}.

{[}24{]} Davoudiasl H, Hooper D, and McDermott SD 2016 \textit{Phys
Rev Lett.} 116 \textbf{031303}.
\end{document}